\documentclass[letter]{emulateapj}

\usepackage{natbib}

\begin{document}

\title{143 GHz brightness measurements of Uranus, Neptune, and
  other secondary calibrators with Bolocam between 2003 and 2010}

\author{J.~Sayers\altaffilmark{1,2}, 
   N.~G.~Czakon\altaffilmark{1},
   and S.~R.~Golwala\altaffilmark{1},
 }
\altaffiltext{1}
  {California Institute of Technology, Pasadena, CA 91125}
\altaffiltext{2}
  {jack@caltech.edu}

\begin{abstract}

Bolocam began collecting science data in 2003 as the long-wavelength
imaging camera at the Caltech Submillimeter Observatory.
The planets, along with a handful of secondary calibrators,
have been used to determine the flux calibration for all of the
data collected with Bolocam.
Uranus and Neptune stand out as the only two planets
that are bright enough to be seen with high signal-to-noise
in short integrations without saturating the standard Bolocam readout
electronics.
By analyzing all of the 143 GHz observations 
made with Bolocam between 2003 and 2010, we find that the brightness
ratio of Uranus to Neptune is $1.027 \pm 0.006$, with no evidence
for any variations over that period.
Including previously published results at $\simeq 150$~GHz,
we find a brightness ratio of
$1.029 \pm 0.006$ with no evidence for time variability
over the period 1983-2010.
Additionally, we find no evidence for time-variability in the
brightness ratio of either Uranus or Neptune to the ultracompact HII
region G34.3 or the protostellar source NGC 2071IR.
Using recently published WMAP results we constrain the absolute
143~GHz brightness of both Uranus and Neptune to $\simeq 3$\%.
Finally, we present $\simeq 3$\% absolute 143~GHz 
peak flux density values for the 
ultracompact HII regions G34.3 and K3-50A and the
protostellar source NGC 2071IR.

\end{abstract}

\section{Introduction}

The past two decades have seen vast improvements in the 
sensitivity and detector counts of broadband (sub)millimeter
imaging cameras,
from single pixel receivers to background-limited kilopixel
arrays (e.g., \citet{duncan90, holland99, runyan03, dowell03, 
haig04, carlstrom09, swetz11, holland06}).
Many of these instruments have relied on the planets, along 
with a handful of secondary calibrators, to obtain
$\lesssim 10$\% flux calibrations 
(e.g., \citet{hill09}, \citet{rudy87}, 
\citet{griffin93}, and \citet{sandell94}).
Alternatively, some of the
large-scale survey instruments have
used cosmic microwave background (CMB) measurements to
calibrate their data to even better precision 
(e.g., \citet{reichardt09}).
However, calibrating from the CMB is only practical for
instruments that image $\gtrsim 100$ deg$^2$,
and therefore cannot be done in many cases.
When planets and secondary calibrators are used,
the accuracy of the flux calibration
is often limited by uncertainties in the brightness of these
calibrators rather than by measurement uncertainties
(e.g., \citet{sayers09}). 
Consequently, accurate measurements of the absolute
brightness of the planets and secondary calibrators,
along with a detailed understanding of any temporal variability
in these brightnesses, is critical to
maximizing the scientific potential of
current and future (sub)millimeter cameras.

Until recently, systematic uncertainties in the
brightness of Uranus and Neptune have been $\simeq 5-10$\%,
owing to uncertainties in the Martian brightness used
to determine the absolute brightness of each of these planets
(\citet{wright76}, \citet{rudy87}, \citet{orton86}, G93).
However, recently published WMAP results have significantly
improved the constraints on the brightness of Mars ($\simeq 1$\%)
and 
Uranus ($\simeq 3$\%)
by calibrating them relative to the CMB
\citep{page03, hill09, weiland11}.
WMAP has also published CMB-calibrated brightness
measurements of Neptune accurate to 8\% \citep{weiland11}.

Additionally, recently published results have shown  
evidence for a 
$\simeq 10$\% decrease in the brightness of 
Uranus at 8.6 and 90~GHz over the 
$\simeq 20$ year period from the mid 1980s
to the mid 2000s (\citet{klein06} and
\citet{kramer08}),
which have been attributed to changes in the
relative viewing angle of the planet from
earth due to its large obliquity of 82 deg.
Note that, over the period of observations in  
Kl06 and Kr08 the SEP latitude of Uranus ranged from
a minimum of $-82$~deg in 1985 to a maximum 
of $-7$~deg in 2005.
These data indicate that the brightness of
Uranus increases as the magnitude of the sub-earth
point (SEP) latitude increases (i.e., the south pole 
of Uranus is brighter than the equatorial 
regions).
Further evidence for this scenario comes from
resolved cm-wave images of Uranus, which
indicate that the poles are brighter than the equatorial
regions \citep{orton07}.

\section{Observations}

Bolocam is a large-format bolometric camera
with an 8~arcmin field of view capable of
observing at either 143 or 268~GHz.
Bolocam began collecting scientific data in early 2003
at the Caltech Submillimeter Observatory (CSO),
and was commissioned as a facility instrument
later that year~\citep{haig04}.
In this manuscript we focus exclusively on 
143~GHz Bolocam data,
which was collected between 2003 
and 2010 during nine separate observing runs
(see Table~\ref{tab:obs}).
All of the data collected with Bolocam has been
calibrated using the planets (generally Uranus and Neptune),
along with sources given in \citet{sandell94}.
The Bolocam flux calibration procedure is
described in detail in
\citet{laurent05} and \citet{sayers09},
and we briefly describe the details below.

\begin{deluxetable}{cccccc} 
  \tablewidth{0pt}
  \tablecaption{Summary of 143~GHz Bolocam observations}
  \tablehead{\colhead{date} & \colhead{Uranus} &
    \colhead{Neptune} & \colhead{NGC 2071IR} &
    \colhead{G34.3} & \colhead{K3-50A}}
  \startdata
    2003/11 & \phn6 & \phn5 & \phn9 & \phn0 & \phn0 \\
    2004/11 & 11 & 20 & \phn0 & \phn0 & \phn0 \\
    2006/11 & \phn6 & \phn3 & 15 & \phn0 & \phn0 \\
    2008/06 & \phn2 & \phn3 & \phn0 & \phn5 & \phn0 \\
    2008/07 & \phn1 & \phn2 & \phn0 & 10 & \phn2 \\
    2008/11 & \phn7 & \phn0 & \phn8 & \phn0 & \phn0 \\
    2009/10 & \phn5 & 21 & \phn9 & \phn0 & \phn0 \\
    2010/02 & \phn0 & \phn0 & 18 & 12 & \phn0 \\
    2010/10 & \phn6 & \phn4 & \phn0 & \phn0 & \phn0

  \enddata
  \tablecomments{The number of integrations for each
    flux calibrator for nine 143~GHz Bolocam observing runs
    between 2003 and 2010.
    Note that there were five additional 143~GHz 
    observing runs during that period, but those
    five observing runs included observations of 
    only one flux calibration source, and are 
    therefore excluded from our analysis.
    Note that the typical S/N of each integration
    is $> 100$.}
  \label{tab:obs}
\end{deluxetable}

Fundamental to our flux calibration technique 
is the fact that Bolocam continuously monitors
the operating resistance of the bolometers via
the bias carrier amplitude.
As the transmission of the atmosphere increases
the optical load on the bolometers decreases
and the operating resistance increases.
Additionally, the responsivity (in nV/Jy) of the bolometers
increases as the bolometer resistance
increases.
Consequently, we are able to simultaneously
account for changes in atmospheric transmission
and detector responsivity by fitting the 
flux calibration as a function of bias carrier
amplitude.
In practice, the technique works as follows.
For a given observing run, we generally observe multiple
sources that have constant brightnesses.
We then simultaneously fit the data for all
of these sources according to
\begin{displaymath}
  V_i(R_{j}) = B_i \Omega_{i,j} (\alpha_1 + \alpha_2 R_{j}), 
\end{displaymath}
where $V_i$ is the bolometer response to source $i$
(in nV), $B_i$ is the brightness of source $i$,
$\Omega_{i,j}$ is the solid angle of source $i$ during
observation $j$,
$R_{j}$ is the bolometer resistance 
during observation $j$ as measured
by the bias carrier amplitude, 
and $\alpha_1$ and $\alpha_2$ describe how the
bolometer response changes as a function of $R_j$.

Implicit in the above formula is the assumption that
changes in the overall opacity of the atmosphere 
are completely accounted for via $\alpha_1$ and $\alpha_2$,
which implies that the atmospheric transmission
varies in a such way that the shape of the
Bolocam bandpass through the atmosphere is constant.
However, over the range of conditions where these data were
collected (column depths of precipitable water
between 0.5 and 3.0~mm), variations in the
atmospheric opacity do cause the Bolocam
bandpass to vary slightly 
\citep{pardo01, pardo01_2, pardo05}).
For a typical observing run, these bandpass variations
in the atmosphere add an additional uncertainty
of $\simeq 0.2$\% to our measured brightness
ratios, which is negligible compared to our
measurement uncertainties of $1-5$\%.

\begin{figure*}
  \centering
  \includegraphics[width=.30\textwidth]{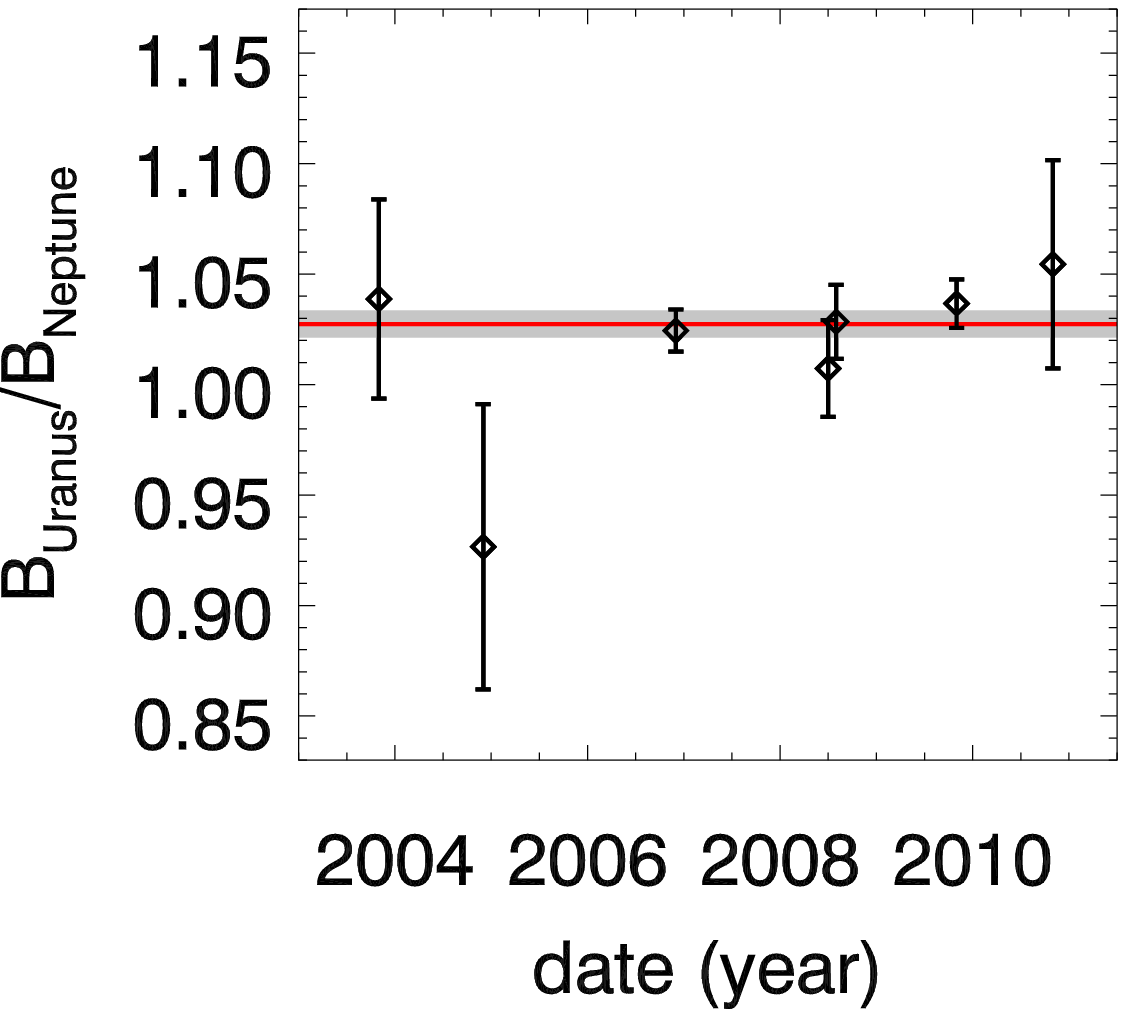}
    \hspace{.04\textwidth}
  \includegraphics[width=.30\textwidth]{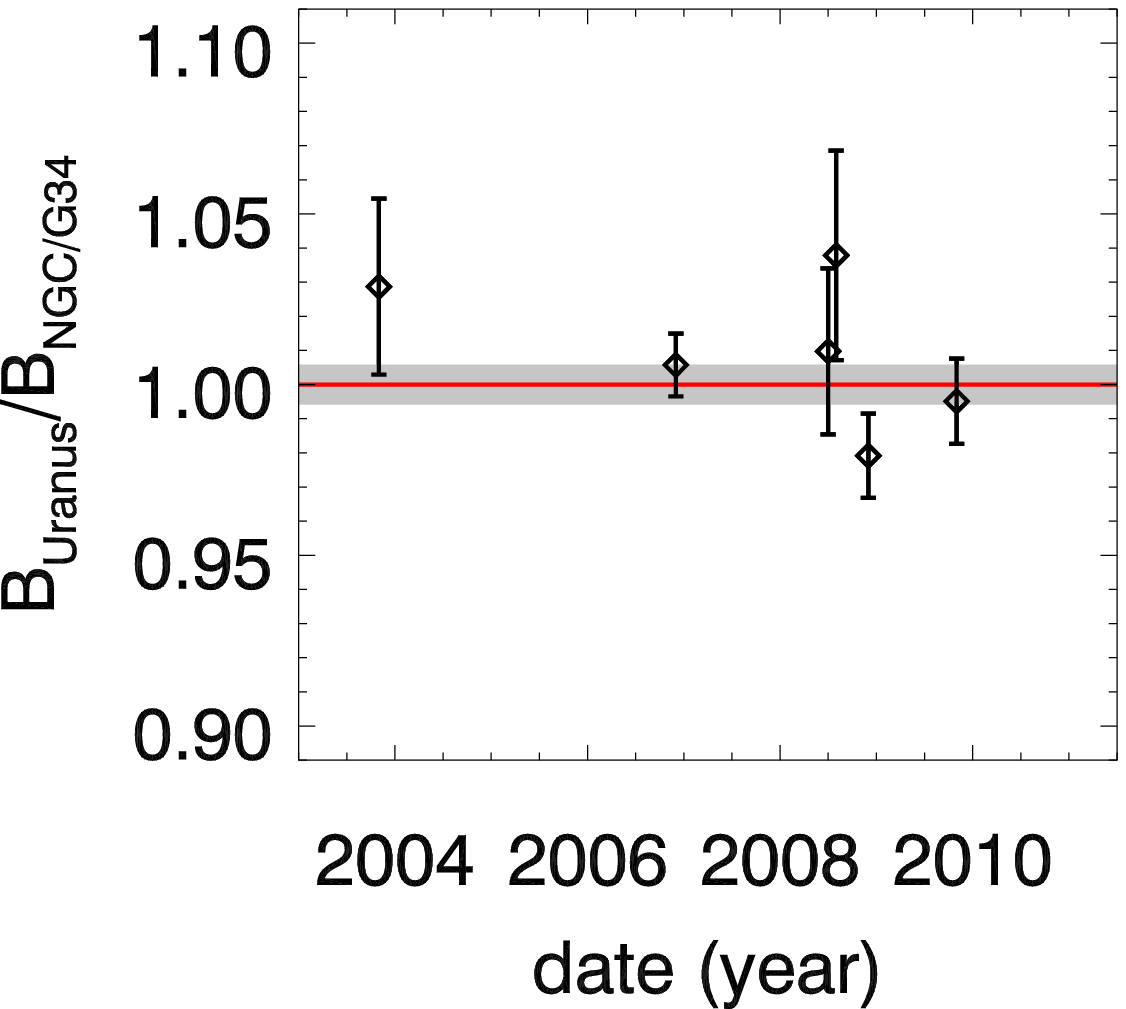}
    \hspace{.04\textwidth}
  \includegraphics[width=.30\textwidth]{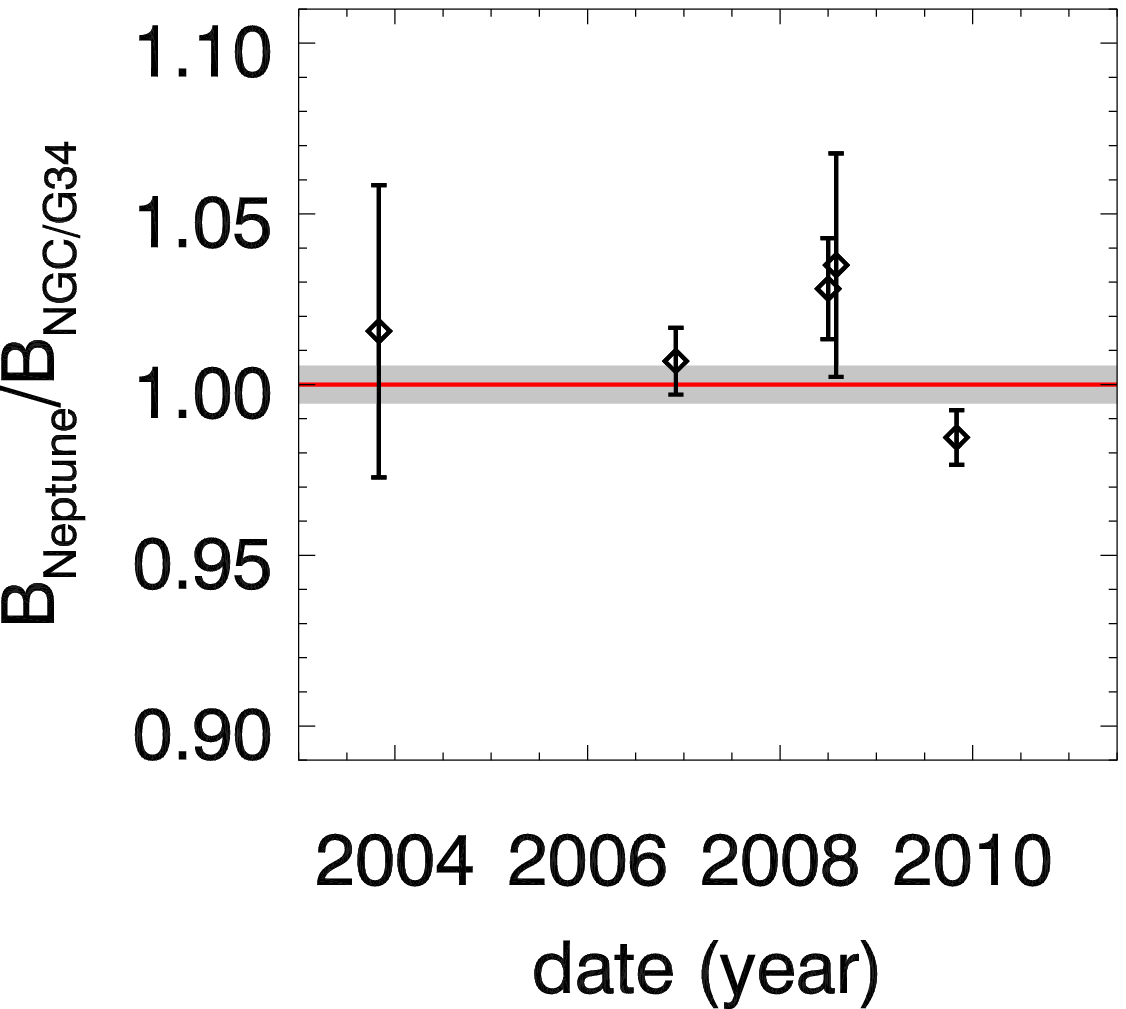}
  \caption{From left to right, Bolocam measurements of the brightness ratio
    of Uranus to Neptune, the relative brightness ratio of Uranus to
    NGC 2071IR or G34.3, and the relative brightness ratio
    of Neptune to NGC 2071IR or G34.3 for the period from 2003 to 2010.
    The red line shows the best-fit mean value of the brightness
    ratio (equal to 1 by definition for the center and right plots),
    and the grey band represents the 1$\sigma$ confidence region.
    Our data show no evidence for temporal variations in any of
    these brightness ratios.}
  \label{fig:bolocam_ratios}
\end{figure*}

Note that we have assumed that the angular size of 
the source is much smaller than the angular size
of the Bolocam point-spread function, which
is true for all of the sources\footnote{
The semi-diameters of the sources in arcsec are
$\simeq 2$ (Uranus), $\simeq 1$ (Neptune),
$\simeq 6$ (NGC 2071IR), $\simeq 3.5$ (G34.3),
and $\simeq 4$ (K3-50A), all of which are
small compared to Bolocam's
59~arcsec full-width at half-maximum
point-spread function \citep{sandell94}.
The secondary calibrators are large enough that
our point-like assumption will cause us to underestimate
the total surface brightness of these objects by
$\lesssim 1$\%
(see Section~\ref{sec:results}).
Uranus and Neptune are small enough that the correction
is negligible ($< 0.1$\%).}.
Since the solid angle of both Uranus and Neptune
varies with observation epoch,
we compute the value of $\Omega_{i,j}$ separately 
for each
integration of these planets
using the James Clerk Maxwell Telescope (JCMT)
Fluxes program\footnote{http://www.jach.hawaii.edu/jac-bin/planetflux.pl}.
For the secondary calibrators, the value of $\Omega_{i,j}$ 
is assumed to be constant.
See Table~\ref{tab:ratio} and Figure~\ref{fig:bolocam_ratios}.

\begin{deluxetable}{cccc} 
  \tablewidth{0pt}
  \tablecaption{Bolocam 143~GHz brightness ratios}
  \tablehead{\colhead{date} & \colhead{$\frac{B_{Uranus}}{B_{Neptune}}$} &
    \colhead{$\frac{B_{Uranus}}{B_{NGC/G34}}$} &
    \colhead{$\frac{B_{Neptune}}{B_{NGC/G34}}$}}
  \startdata
    2003/11 & $1.039 \pm 0.045$ & $1.029 \pm 0.026$ & $1.016 \pm 0.042$ \\ 
    2004/11 & $0.927 \pm 0.065$ & - & - \\ 
    2006/11 & $1.024 \pm 0.010$ & $1.006 \pm 0.009$ & $1.007 \pm 0.010$ \\ 
    2008/06 & $1.007 \pm 0.022$ & $1.010 \pm 0.024$ & $1.028 \pm 0.015$ \\ 
    2008/07 & $1.028 \pm 0.017$ & $1.038 \pm 0.030$ & $1.035 \pm 0.032$ \\ 
    2008/11 & - & $0.979 \pm 0.012$ & - \\ 
    2009/10 & $1.037 \pm 0.011$ & $0.995 \pm 0.012$ & $0.985 \pm 0.008$ \\ 
    2010/02 & - & - & - \\ 
    2010/10 & $1.054 \pm 0.047$ & - & - \\ 
    \bf{total} & \bf{$1.027 \pm 0.006$} & \bf{$1.000 \pm 0.006$} & \bf{$1.000 \pm 0.006$}
  \enddata
  \tablecomments{Brightness ratios of Uranus to Neptune,
    Uranus to the secondary calibrators NGC 2071IR and G34.3,
    and Neptune to the same secondary calibrators.
    Note that the brightness ratios of the planets to
    the secondary calibrators are normalized
    to the mean brightness ratio since the
    solid angle of the secondary calibrators is
    not precisely known.
    The brightness ratio of NGC 2071IR to G34.3 is constrained
    using the 2010/02 data.}
  \label{tab:ratio}
\end{deluxetable}

\section{Results}
\label{sec:results}

Seven Bolocam observing runs contain observations of both
Uranus and Neptune.
In each run, we measure the brightness ratio of
Uranus to Neptune with $\lesssim 5$\% statistical
uncertainty.
If we assume that the brightness ratio is
constant over the period from 2003 to 2010, then
we find that $B_{Uranus}/B_{Neptune} = 1.027 \pm 0.006$
with $\chi^2/\textrm{DOF} = 4.5/6$.
A linear fit versus year to the data reduces the value of $\chi^2$
by 0.9, indicating that a linear fit is not
required (a F-test shows that 39\% of
realizations of constant data would have yielded
a larger F-ratio, implying that
a constant fit to our data is adequate).
Comparing to published results at similar frequencies,
O86 find a brightness ratio of $1.016 \pm 0.038$
using 150~GHz data collected between February 1983 and March 1984 at 
the National Radio Astronomy Observatory 12-m antenna at Kitt
Peak,
and G93 find a brightness ratio of $1.043 \pm 0.016$
using 156~GHz data collected at the 
JCMT in May 1990/1992.
The combined $\simeq 150$~GHz data set of O86, G93, and Bolocam
is well described by
a constant brightness ratio of $B_{Uranus}/B_{Neptune} = 1.029 \pm 0.006$,
with a $\chi^2/\textrm{DOF} = 5.4/8$.
In this case a linear fit versus year reduces the value of $\chi^2$ by
0.1, and a F-test shows that 76\% of realizations
of constant data would have produced a larger F-ratio.
In addition, the linear fit provides a formal 95\% 
confidence level upper limit of 5.3\% on the
magnitude of the variation over the
28-year period 1983 to 2010
(SEP latitudes of $-82$ to $+10$).

In addition to searching for variations in the brightness
ratio of Uranus and Neptune, we also 
searched for temporal variations
in the brightness of each planet relative to sources
that
are known to have constant brightnesses, NGC 2071IR
and G34.3 \citep{sandell94}.
Six observing runs contained observations of Uranus and either
NGC 2071IR or G34.3, and five observing runs contained
observations of Neptune and either NGC 2071IR or G34.3.
Note that we constrained the brightness ratio of NGC 2071IR
and G34.3 based on observations of both objects
in February 2010.
A constant fit of the brightness ratios yields
a $\chi^2$/DOF$ = 6.3/5$ for Uranus and 
a $\chi^2$/DOF$ = 9.1/4$ for Neptune. 
We again calculate the F-ratio for a constant fit to the 
brightness ratios compared to a linear fit versus year
and again find that there is not compelling evidence
for a linear fit
(22\% of realizations of constant data would have a larger
F-ratio than the Uranus data and
21\% of realizations of constant data would have a larger
F-ratio than Neptune data).

Since we are unable to measure any temporal variations
in the brightnesses of Uranus or Neptune, we use the
WMAP measurements of the brightness temperature of
each planet~\citep{weiland11}, along with our measurement
of the brightness ratio (1.027) to determine the
143~GHz brightness temperature of each planet.
In G93 the authors present
models for the temperature spectra
of each planet based on observations between
90 and 850~GHz, and we have used these models
to extrapolate the 94~GHz WMAP measurements
to 143~GHz.
G93 state that these models are
accurate to $<2$~K when absolute calibration
uncertainties are not included, and we have
therefore included a 2~K uncertainty
in the model-based extrapolation of the 
absolute 94~GHz WMAP measurement to 143~GHz.
We find that the brightness temperature of Uranus is
$106.6 \pm 3.5$~K and the brightness temperature of Neptune is
$103.8 \pm 3.4$~K, where the uncertainties include the WMAP
and Bolocam measurement uncertainties, along with
the quoted uncertainties in the G93 models.
These brightness temperatures indicate that the 
G93 models need to be scaled by
$0.931 \pm 0.031$ (Uranus) and $0.946 \pm 0.031$ (Neptune).
Since the G93 brightness models of
Uranus and Neptune were normalized to
the Martian calibration model described in 
\citet{griffin86} (based on results from
\citet{wright76} and \citet{ulich81}),
our results indicate that the \citet{griffin86}
model over-predicts the brightness of Mars
by $\simeq 5-7$\%.
These scalings are consistent with the results
of \citet{halverson09}, who find that the \citet{wright76}
model should be scaled
by a factor of $0.922 \pm 0.037$ based on the 
WMAP Martian brightness measurements given in \citet{hill09}.

\begin{deluxetable}{ll} 
  \tablewidth{.75\columnwidth}
  \tablecaption{Absolute 143~GHz Brightness Values}
  \tablehead{\colhead{object} & \colhead{brightness}}
  \startdata
    Uranus & $106.6 \pm 3.5$~K \\
    Neptune & $103.8 \pm 3.4$~K \\
    NGC 2071IR & $1.315 \pm 0.043$ Jy\\
    G34.3 & $13.87 \pm 0.47$ Jy \\
    K3-50A & $7.387 \pm 0.244$ Jy
  \enddata
  \tablecomments{Absolute 143~GHz brightness
    values for calibrators observed by Bolocam.
    Uranus and Neptune are given in surface 
    brightness units (K) and the secondary
    calibrators are given in units of peak flux
    density per beam (Jy).}
  \label{tab:cals}
\end{deluxetable}

Using these absolute brightness temperatures
of Uranus and Neptune,
we find 
peak flux densities per beam of $1.315 \pm 0.006 \pm 0.043$~Jy
(NGC 2071IR), 
$13.87 \pm 0.10 \pm 0.46$~Jy (G34.3), and 
$7.387 \pm 0.009 \pm 0.244$~Jy (K3-50A)
for the three secondary calibrators
(see Table~\ref{tab:cals}).
The first value represents our measurement uncertainty
on each peak flux density, while the second value
represents our 3.3\% uncertainty in the brightness 
temperatures of the planets.
Given the source sizes in \citet{sandell94},
along with Bolocam's 59~arcsec FWHM PSF,
these values should be within 1\% of the total
flux density for each source.
Note that \citet{sandell94} found the following peak flux
densities per 27~arcsec beam for these sources at 150~GHz:
$1.5 \pm 0.2$~Jy (NGC 2071IR), $12.5 \pm 0.4$~Jy (G34.3), and 
$6.5 \pm 0.2$~Jy (K3-50A), where the uncertainties do not 
include systematic uncertainties in the planet models
used to calibrate the data.
Our measured peak flux densities are $\simeq 10$\% higher
for both G34.3 and K3-50A
(plus an additional factor of $\simeq 7$\% due to
the fact that \citet{sandell94}
used the G93 model in order
to determine his absolute calibration,
which we have found to overestimate
the brightness of Uranus by $\simeq 7$\%).
It is not clear why our measured peak flux densities are
higher for G34.3 and K3-50A,
although the known extended emission in both sources,
coupled with the fact that \citet{sandell94}
made single-pixel chopped photometry
measurements at 27~arcsec resolution,
may be the cause.

\section{Discussion}

\begin{figure}
  \includegraphics[width=\columnwidth]{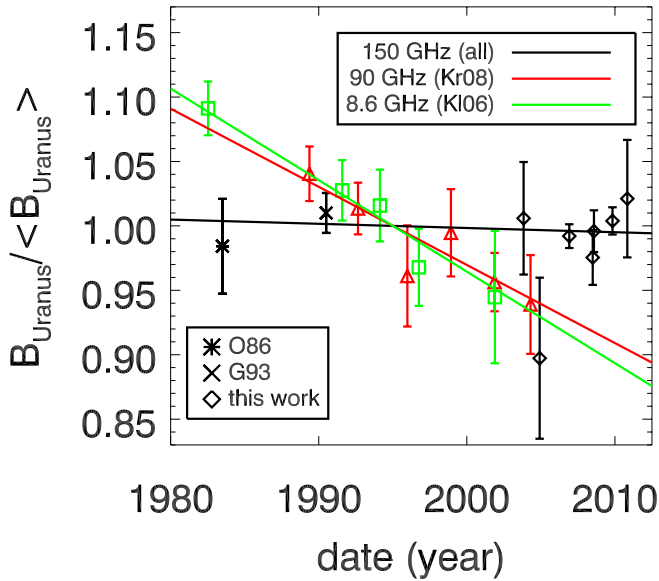}
  \caption{Uranus brightness measurements relative to the
    average brightness of Uranus at
    a range of frequencies between 1980 and 2010.
    The black points show $\simeq 150$~GHz data from
    O86, G93, and Bolocam, the red triangles show
    90~GHz data from Kr08, and the green squares
    show 8.6~GHz data from Kl06 (taken from
    Table 2 of each manuscript
    and rebinned for visualization).}
  \label{fig:f2}
\end{figure}

We find that Uranus and Neptune behave as ideal
sources for flux calibration at 143~GHz, with
no evidence for temporal brightness variations.
For Uranus, these results are in contrast to 
the lower frequency measurements of
Kl06 and Kr08,
who find $\simeq 0.5$ percent/year variations
in the brightness temperature of Uranus at 8.6 and 90~GHz.
Our data, combined with the measurements
of O86 and G93, place a 95\% confidence level upper
limit of 0.19 percent/year 
on the magnitude of variations in the brightness
temperature of Uranus at $\simeq 150$~GHz over the same 
period.
See Figure~\ref{fig:f2}.
A physical interpretation of the temporal
variations in the brightness of Uranus seen at lower
frequencies by
Kl06 and Kr08,
in combination with our static 143~GHz results, 
is beyond the scope
of this manuscript, which is intended to quantify
the magnitude and stability of the brightness
of Uranus for the purposes of using it as a 
calibrator at 143~GHz.
However, we do note that the combined
results are not necessarily inconsistent, given that
higher frequency observations
of Uranus probe higher altitudes in the atmosphere (Kr08).

\section{Conclusions}

Using Bolocam data collected between 2003 and 2010
we have tightly constrained the 143~GHz brightness ratio
of Uranus and Neptune ($1.027 \pm 0.006$), and we find no
evidence for temporal variations in the 143~GHz brightness
temperature of either planet over that period.
Combining our results with those of O86 and G93,
we find no evidence for 143~GHz brightness variations
in either planet over the period from $1983-2010$,
and place a 95\% confidence level upper limit of 5.3\%
on the magnitude of brightness variations over the 28
year period from 1983 to 2010.
By extrapolating the WMAP 94~GHz results given in 
\citet{weiland11} to our observing band
using the brightness models
presented in G93, we are able to constrain the absolute
143~GHz brightness temperature of each planet to $\simeq 3$\%.
Additionally, we determine $\simeq 3$\% absolute 
143~GHz peak flux densities for the 
ultracompact HII regions G34.3 and K3-50A and the
protostellar source NGC 2071IR.

\section{Acknowledgments}

We acknowledge the assistance of: 
the Bolocam instrument team:
Peter Ade, James Aguirre, James Bock, Sam Edgington,
Jason Glenn, Alexy Goldin, Sunil Golwala, 
Douglas Haig, Andrew Lange, Glenn Laurent,
Phil Mauskopf, Hien Nguyen, Philippe Rossinot, and Jack Sayers;
Matt Ferry, Matt Hollister, Patrick Koch, Kai-Yang Lin, Sandor Molnar, Seth Siegel,
and Keiichi Umetsu who, in addition to
the Bolocam instrument team, helped collect the data 
presented in this manuscript;
the day crew and Hilo
staff of the Caltech Submillimeter Observatory, who provided
invaluable assistance during commissioning and data-taking for this
data set; 
Kathy Deniston, Barbara Wertz, and Diana Bisel, who provided effective
administrative support at Caltech and in Hilo;
and the referee, who provided helpful comments and suggestions.  
Bolocam was constructed and
commissioned using funds from NSF/AST-9618798, NSF/AST-0098737,
NSF/AST-9980846, NSF/AST-0229008, and NSF/AST-0206158.  JS 
was partially supported by a
NASA Graduate Student Research Fellowship, a NASA
Postdoctoral Program fellowship, NSF/AST-0838261, 
and NASA/NNX11AB07G;
NC was partially supported by NASA Graduate Student
Research Fellowship;
SG acknowledges an Alfred P. Sloan Foundation fellowship.

\end{document}